\documentclass[12pt]{article}
\def\be{\begin{equation}} \def\ee{\end{equation}} \def\bea{\begin{eqnarray}}
\def\eea{\end{eqnarray}}

\usepackage[dvips]{graphicx}
\usepackage{dcolumn}
\usepackage{bm}

\begin{document}

\renewcommand{\thefootnote}{\fnsymbol{footnote}}
\setcounter{footnote}{0}

\begin{center}
\vskip 5mm 
{\Large\bf 
Neutrino Magnetic Moment Contribution to the Neutrino-Deuteron Reaction
}
\vskip 5mm 
{\large 
 K. Tsuji$^{a}$,  S. Nakamura$^{a,b}$,  T. Sato$^{a,b}$,\\
  K. Kubodera$^{b}$,  and F. Myhrer$^{b}$\footnote{
E-mail address:tsuji@kern.phys.sci.osaka-u.ac.jp(K. Tsuji),
tsato@phys.sci.osaka-u.ac.jp(T. Sato),
nakkan@rcnp.osaka-u.ac.jp(S. Nakamura),
kubodera@sc.edu(K. Kubodera),
myhrer@sc.edu(F. Myhrer)}
}

\vskip 5mm

{\it
$^{a}$ Department of Physics, Osaka University, Toyonaka, Osaka 560-0043, 
      Japan \\
 $^{b}$ Department of Physics and Astronomy, University of South Carolina,
      Columbia,\\ SC 29208, USA
}
\vskip 5mm

\end{center}

\vskip 8mm

We study the effect of the neutrino magnetic moment
on the neutrino-deuteron breakup reaction,
using a method called 
the standard nuclear physics approach,
which has already been well tested for several
electroweak processes involving the deuteron.


\newpage
\renewcommand{\thefootnote}{\arabic{footnote}}
\setcounter{footnote}{0}

The Sudbury Neutrino Observatory (SNO)
has given precision data on the flux
of the $^8$B solar neutrinos
and on its flavor content,
and these data have provided solid evidence 
for neutrino oscillations \cite{sno,snox}. 
The SNO experiments measure
the yield from the charged current (CC) reaction, 
$\nu_e + d \rightarrow e^- + p + p$,
which occurs only for the electron neutrino,
and the yield from the neutral current (NC) reaction, 
$\nu_x + d \rightarrow \nu_x + p + n$\,\,
($x=e,\,\mu,\tau$), 
which takes place 
with the same cross section
for any neutrino flavor $x$. 
In interpreting the SNO data 
the theoretical estimates
of the $\nu$-$d$ cross sections play an important role.
Detailed studies of the $\nu$-$d$ reaction 
in the solar neutrino energy region have been
carried out using the 
standard nuclear physics approach (SNPA)
\cite{nsgk,netal},
pion-less effective field theory based
on the power divergence subtraction scheme 
\cite{butler1,butler2}, 
and EFT* (or MEEFT) \cite{ando},
which is a ``pragmatic" version of 
chiral perturbation theory in the Weinberg counting scheme
\cite{Parketal03}. 
Apart from the radiative corrections,  
the uncertainties in the calculated values of  
the $\nu$-$d$ cross sections
are estimated to be 1\%, see Ref.\cite{netal}.

\vspace{3mm} 
The above-mentioned calculations
use the weak interaction Hamiltonian
dictated by the standard model,
in which the neutrinos interact with the deuteron 
only by exchanging the $W$ and $Z$ bosons.
Meanwhile, if the neutrino has a magnetic moment
(let it be denoted by $\mu_\nu$), 
it can interact with the deuteron 
via a photon exchange as well.
This additional interaction,
contributing to the 
$\nu_x d \rightarrow \nu_x p n$ reaction,
can affect the interpretation of 
the NC data of SNO at some level of precision.
In the following,
the electromagnetic neutrino-deuteron
interaction due to $\mu_\nu$
shall be simply referred to as 
the $\mu_\nu$-interaction.
Since the contribution of the $\mu_\nu$-interaction
to the $\nu_x d \rightarrow \nu_x p n$ reaction
and the NC contribution do not interfere with each other 
(see below),
we can decompose the cross section as 
\begin{eqnarray}
\sigma(\nu_x d\rightarrow \nu_x pn)
=\sigma_{EM} + \sigma_{NC},\label{eq:decompose}
\end{eqnarray} 
where $\sigma_{EM}$ stands for the contribution of
the $\mu_\nu$-interaction.
Akhmedov and Berezin (AB) \cite{ab} estimated $\sigma_{EM}$,
using the effective range approximation
and the impulse approximation. 
Recently, Grifols, Masso and Mohanty (GMM) 
\cite{mohanty}
gave a new estimate of the $\mu_\nu$-interaction effect
based on the equivalent photon approximation.
GMM reported a significant difference
between their value of $\sigma_{EM}$ 
and the value obtained by AB \cite{ab}, 
but the origin of this discrepancy
was left undiscussed.

\vspace{3mm}
We carry out here a calculation of 
$\sigma_{EM}$ which is free from
the effective-range approximation, 
the impulse approximation and the 
equivalent photon approximation.
To this end, we make an appropriate 
extension of SNPA used in Refs.\cite{nsgk,netal}
and calculate $\sigma_{EM}$, 
including both the impulse and exchange-current terms,
and with the use of the two-nucleon wave functions
generated from a high-precision realistic $NN$ potential. 
We shall also briefly discuss some consequences 
of our calculation in the context
of placing upper limits to the
neutrino magnetic moments.

\vspace{3mm}
We are concerned here with the reaction
\begin{eqnarray}
\nu(k)_a + d(P) \rightarrow \nu(k')_b + p(p_1') + n(p_2')\;   
\label{eq:reaction} 
\end{eqnarray}
at low energies,
where $a$ and $b$ stand for neutrino flavors. 
The relevant transition amplitude $t_{fi}$
consists of two terms,
one arising from the standard NC interaction
and the other from the $\mu_\nu$-interaction. 
Thus
\begin{eqnarray}
t_{fi}  & = & -\frac{G_F}{\sqrt{2}}
j_{\mu }^{NC}l^{NC,\mu }
 +\frac{1}{q^2}j_{\mu }^{EM} l^{EM,\mu}, 
\label{t-mat}
\end{eqnarray} 
where $q_{\mu} = k_\mu - k'_\mu$.
The lepton neutral current matrix element, $l^{NC, \mu}$, 
is diagonal in the weak-flavor eigen-states and is given as 
\begin{eqnarray}
l^{\mu } & =&  \bar{u}_b(k')\gamma^\mu 
( 1 - \gamma_5)u_a(k)\delta_{b,a} \; , 
\label{lept-nc} 
\end{eqnarray}
whereas the neutrino electromagnetic current
is given in terms of the magnetic dipole moment 
$\mu_{ba}$ and
the electric dipole moment $\epsilon_{ba}$, 
\begin{eqnarray}
  l^{EM,\mu } & = & \bar{u}_b(k')i\sigma ^{\mu \nu}q_{\nu}
    ( \mu _{ba} - \epsilon _{ba}\gamma _5 )u_a(k) \; ,   
\label{lept-em}
\end{eqnarray}
where $b$ and $a$ stand for the mass eigen-states 
of the neutrinos.
In Eq.(\ref{t-mat}), $j_\mu^{NC}$ and $j_\mu^{EM}$
represent the nuclear matrix matrix elements:
\begin{eqnarray}
j_\mu^{X} & = & <\!p(p_1')n(p_2')| J^{X}_\mu|d(P)\!>
\;\;\;(X=NC,EM)\,,
\end{eqnarray}
where $J^{NC}_\mu$ is the weak NC current, and 
$J^{EM}_\mu$ is the electromagnetic current.
The explicit forms of $J^{NC}_\mu$ and $J^{EM}_\mu$
were described in detail in Refs. \cite{nsgk,netal}. 
At the solar neutrino energies it is safe to apply
non-relativistic approximation to
the one-nucleon (impulse approximation) current. 
The two-nucleon exchange current is derived 
using the one-meson-exchange SNPA model \cite{nsgk,netal}. 
We remark that this SNPA model describes very well the 
$n + p \rightarrow d + \gamma$ reaction 
at low energies,
and that, for $E_n< 0.1$ MeV, 
the magnetic dipole (M1) transition dominates,
whereas for higher energies the
electric dipole (E1) transition dominates; 
see Ref. \cite{wein95} for details. 

\vspace{3mm}
If the neutrino mass is neglected 
in Eqs.(\ref{lept-nc}) and (\ref{lept-em}), 
then $l^{NC, \mu}$ conserves the neutrino helicity 
while $l^{EM,\mu }$ flips the helicity. 
This means that to a very good approximation
there is no interference between the NC and EM contributions,
and therefore the total cross section for
the $\nu_x d \rightarrow \nu_x pn$ reaction
can be written as the incoherent sum of the NC and EM
terms as in Eq.(\ref{eq:decompose}).
An explicit expression for $\sigma_{NC}$
and its elaborate numerical calculation 
can be found in Ref.\cite{nsgk};
our goal here is to evaluate $\sigma_{EM}$
to a comparable degree of accuracy.
It is convenient to express $\sigma_{EM}$ as an 
integral over the final two-nucleon relative kinetic energy,  
$T_{NN} = {\bm{p}'}^2/M$, where 
$\bm{p}'=(\bm{p}'_1 - \bm{p}'_2)/2$,  
\begin{eqnarray}
\sigma_{EM} = \int_0^{T_{max}}dT_{NN} \; 
\frac{d\sigma_{EM}}{dT_{NN}}\, .
\label{eq:sigmaEM}
\end{eqnarray}
Here
\begin{eqnarray}
\frac{d\sigma_{EM}}{dT_{NN}}
 & = & (\mu_{\nu,a})^2 \int_{-1}^1 d\cos\theta
\sum_{L_f,S_f,J_f} \frac{4 \alpha m_N \; 
p' {k'}^3 k}{3 (-q_\mu^2)}
 \Bigg[\,  \frac{(k+k')^2(1-\cos\theta)^2}{\bm{q}^4}
  \sum _{J \ge 0} \big|\!<\!T^J_C\!> \!\big|^2 
  \nonumber \\ & &
  +\frac{\sin^2\theta}{2 \bm{q}^2} \sum _{J \ge 1}
\big\{ \big|\!<\!T^J_M\!>\!\big|^2 
+\big|\! <\!T^J_E\!> \!\big|^2 \big\}\, 
\Bigg] \; , \label{differential}
\label{si}
\end{eqnarray} 
where $\bm{q}$ is the momentum transferred to the 
two-nucleon system;
the neutrino scattering angle $\theta$ is defined by  
$\bm{k}\!\cdot\! \bm{k}' = k k' \cos\theta $ 
in the $\nu$-$d$ center-of-mass system.
$\mu_{\nu a}$ is the {\it effective} neutrino magnetic moment
for the incoming neutrino of flavor $a$ \cite{raf,fy} 
\begin{eqnarray}
 (\mu_{\nu a})^2 & = & \sum_b |\mu_{ba} 
+ \epsilon_{ba}|^2 \; . 
\end{eqnarray} 
In Eq.(\ref{differential}), 
$<\!T^J_X\!>$ represents the nuclear reduced matrix element
of a multipole operator $T^J_X$ of rank $J$:
\begin{eqnarray}
<\!T^J_X\!> & = <\!(L_f,S_f)J_f||T^J_X||d;J_i=1> \; , 
\end{eqnarray} 
where $X=C,\,M$ and $E$ correspond to  
the Coulomb, magnetic and electric multipole operators, 
respectively.    
The explicit expressions for $<\!T^J_X\!>$
can be found in Eqs. (57) and (38)-(40) 
in Ref. \cite{nsgk}. 
In calculating $\sigma_{EM}$ 
at the solar neutrino energies, 
we need to pay particular attention to 
the sizeable final two-nucleon interactions. 
Here, as in Ref.\cite{nsgk},
we use the final two-nucleon distorted S- and P-waves
obtained with the use of the Argonne V18 potential 
\cite{anl}.  

\vspace{3mm}
In order to illustrate 
the possible $\mu_\nu$-interaction effects 
on $\sigma(\nu_x d\rightarrow \nu_x pn)$, 
we need to specify a ``typical" value of $\mu_{\nu a}$.
The radiative corrections within the standard model
leads to an estimate
$\mu_\nu \sim 3 \times 10^{-19} m_\nu/1{\rm eV} [\mu_B]$, 
where $\mu_B$ is the Bohr magneton
\cite{raf,fy};
the value of $\mu_\nu$ larger than this estimate
would be an indication of new physics. 
The current upper limits obtained from 
reactor neutrino experiments are  
$\mu_{\bar{\nu}_e}< 1.0 \times 10^{-10}\mu_B$ 
\cite{MUNU} 
and 
$\mu_{\bar{\nu}_e}<1.3 \times 10^{-10}\mu_B$ 
\cite{TEXONO}.
The limit found using 
the $\nu_\mu$ and $\bar{\nu}_\mu$ 
from $\pi^+$ and  $\mu^+$ decays is 
$\mu_\mu< 6.8 \times 10^{-10}\mu_B$ \cite{LSND}. 
Meanwhile, the limit for the tau neutrino is reported to be
$\mu_\tau< 3.9 
\times 10^{-7}\mu_B$\cite{DONUT}.\footnote{
When neutrino oscillations are taken into account,
a limit deduced from scattering experiments 
should be understood as a limit on 
$\mu_f = \sum_j |\sum_k U_{fk}\mu_{jk} e^{-iE_k L}|^2$,
where $f$ is the flavor of the beam neutrino, 
$k$ and $j$ denote mass eigenstates, and 
$U$ is the mixing matrix \cite{beacom}.}
As for the astrophysical constraints,
it is known that the avoidance of  
excessive stellar cooling requires
$\mu_{\nu a}<(0.3-1)\times 10^{-11}\mu_B$
(see, e.g., Ref.\cite{fy}).
For the sake of definiteness,
we use here $\mu_{\nu a}=10^{-10}\mu_B$
as a representative value.
This completes the specification of all the quantities
appearing in Eqs.(\ref{eq:sigmaEM}), (\ref{differential}).

\begin{figure}[h]
\caption{The cross section, $\sigma_{EM}$, for the
$\nu_x d \rightarrow \nu_x p n$ reaction
due to the $\mu_\nu$-interaction.
$\sigma_{EM}$ has been calculated 
for $\mu_{eff} = 10^{-10}\mu_B$,
and is given in units of ${\rm cm}^2$.}
\includegraphics[width=8cm] {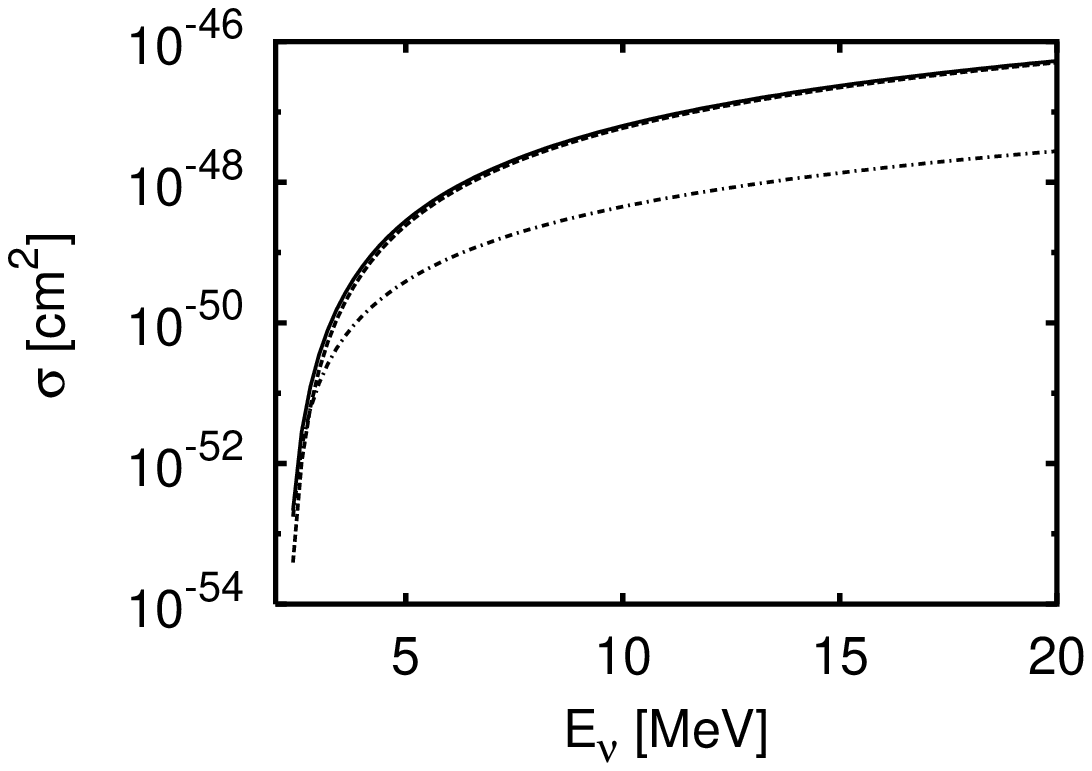} 
\label{fig-dis}
\end{figure}

\vspace{3mm}
Our numerical results for $\sigma_{EM}$ are shown
in table 1 and Fig. \ref{fig-dis}. 
In  Fig. \ref{fig-dis}, 
the dotted curve represents the contribution of 
the Coulomb and E1 transitions
leading to the $^3P_J$ final two-nucleon state,
while the dash-dotted curve gives the contribution of 
the M1 transition leading to the $^1S_0$ state. 
The solid curve shows the value of $\sigma_{EM}$ 
obtained by adding the contributions 
of all the multipoles in Eq.(\ref{si}).
We note that,  
except for the lowest energy region 
($E_\nu < 2.8$ MeV),
$\sigma_{EM}$ is dominated 
by the Coulomb and E1 transitions 
to the final P-wave states. 
This should be contrasted with the case of the NC reaction, 
where the Gamow-Teller transition to the final $^1S_0$ state 
gives a dominant amplitude. 

\begin{figure}[h]
\caption{ 
The ratio $R\equiv \sigma_{EM}/\sigma_{NC}$ 
calculated for $\mu_{eff}=10^{-10}\mu_B$ 
as a function of the incoming neutrino energy. }
\includegraphics[width=8cm] {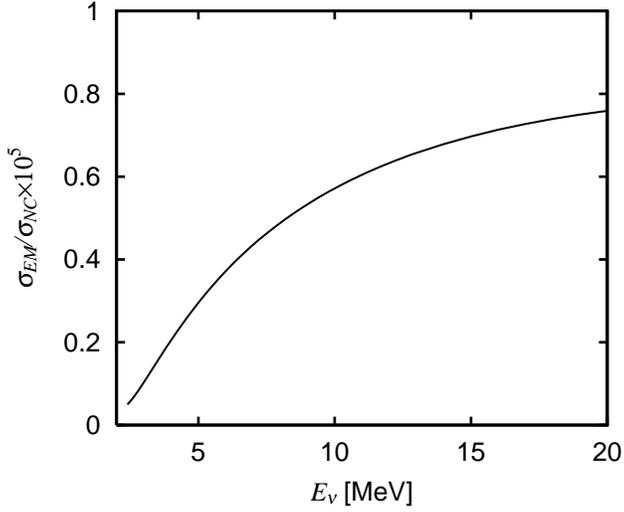}
\label{fig-ratio}
\end{figure}

\vspace{3mm}
Fig. \ref{fig-ratio} 
shows the ratio  $R\equiv \sigma_{EM}/\sigma_{NC}$ 
calculated from $\sigma_{EM}$ obtained
in the present work and $\sigma_{NC}$ given 
in Ref. \cite{netal}.
$R$ increases with the neutrino energy $E_\nu$ and
reaches $\sim 5 \times 10^{-6}$ at $E_\nu=10$ MeV,
which is however still very small.
The smallness of $R$ reflects the fact 
that in the $\nu$-$d$ disintegration reaction, 
the NC contribution comes from an allowed transition, 
while the EM contribution 
arises from a `first-forbidden' transition. 
The latter is intrinsically suppressed 
by a small factor $q\!<\!r\!>$,
where $<\!r\!>$ is a typical nuclear size.
We remark that the contribution 
of the $\mu_\nu$-interaction is much larger for 
elastic $\nu$-$d$ scattering 
than for the $\nu$-$d$ breakup reaction,
since the elastic scattering can occur 
via an allowed-type transition;
it is unfortunate that $\nu$-$d$ elastic scattering
is at present practically impossible to detect.

\vspace{3mm}
To facilitate comparison of our present results
with those in the literature,
we let $\sigma_{EM}$(TNSKM), 
$\sigma_{EM}$(AB), $\sigma_{EM}$(GMM)
denote the values of $\sigma_{EM}$
obtained by us, by AB \cite{ab} and by GMM \cite{mohanty},
respectively.
To compare $\sigma_{EM}$(TNSKM) and
$\sigma_{EM}$(AB),
we read off the numerical value of $\sigma_{EM}$(AB)
from  Fig.1 in Ref.\cite{ab}
and normalize it to the case of 
$\mu_\nu = 10^{-10}\mu_B$, 
the value used in our present calculation.
Comparison after this normalization
indicates that 
$\sigma_{EM}$(TNSKM) is about 4 times 
larger than $\sigma_{EM}$(AB).
As mentioned, our present calculation based on SNPA 
is free from the effective-range approximation 
and the impulse approximation, 
whereas the calculation in Ref.\cite{ab}
does involve these two approximations.
However, this difference in formalisms {\it per se}
is not expected to lead to
a discrepancy as large as a factor of $\sim 4$.
For further comparison,
we find it illuminating to simplify our full calculation
by introducing the effective-range approximation
and impulse approximation;
the resulting simplified formula 
for the E1 contribution
is found to be larger than the E1 part 
of Eq.(18) in Ref.\cite{ab}
by a factor of 2.
It is likely that the remaining discrepancy 
by a factor of $\sim$2 is   
of the numerical origin.
We have done a little numerical exercise
of calculating ``our version" of $\sigma_{EM}$(AB)
directly from Eqs.(18) and (A1) of AB.
Let $\sigma_{EM}$(AB;direct) stand for the result
of this exercise;
the value read off from Fig.1 in AB
is denoted by $\sigma_{EM}$(AB;figure).
If we use $\sigma_{EM}$(AB;direct)
instead of $\sigma_{EM}$(AB;figure),
then the difference between our result and that of AB
becomes a factor of $\sim$2 (instead of $\sim$4),
which is consistent with the above-mentioned difference
in the analytic expressions.

\vspace{3mm}
As far as comparison with GMM \cite{mohanty} is concerned,
we first discuss $<\!\!\sigma_{EM}\!\!>$
defined as 
\begin{eqnarray}
<\!\!\sigma_{\alpha}\!\!> & = &
  \int dE_\nu f(E_\nu) \sigma_{\alpha}(E_\nu),
\end{eqnarray}
where $\alpha=EM, NC$ and $f(E_\nu)$ is
the normalized $^8$B neutrino spectrum.
If we use the value given in GMM, we find 
$<\!\!\sigma_{EM}{\rm (TNSKM)}\!\!>
\,\approx\, <\!\!\sigma_{EM}{\rm (GMM)}\!\!>$,
but that the former is smaller 
than the latter by about 10 \%.
To gain some insight into this difference,
we have examined the analytic
expression for the cross section in GMM
and have confirmed that Eq.(16) in GMM
is indeed valid in the long wavelength approximation
and in a regime where the dominance of the electric 
and Coulomb dipole transitions holds.  
Furthermore, if we calculate $\sigma_{EM}$
directly from Eqs.(16) and (17) of GMM,
the result agrees with $\sigma_{EM}{\rm (TNSKM)}$
within 1\%. 
Therefore, the above-mentioned $\sim$10 \% discrepancy
does not seem to be due to 
the equivalent-photon approximation 
used by GMM.  We have not been able to identify
the origin of the $\sim$10 \% difference
found in the comparison of 
$<\!\!\sigma_{EM}\!\!>$.\footnote{
In GMM's article \cite{mohanty},
there is the statement that 
$\sigma_{EM}$(GMM) is smaller than 
$\sigma_{EM}$(AB) by a factor of $\sim2$.
A comment is in order on this point.
As mentioned, $\sigma_{EM}$(AB) in fact
can represent either $\sigma_{EM}$(AB;figure)
or $\sigma_{EM}$(AB;direct).
In the former case,
we are led to 
$\sigma_{EM}$(TNSKM)
$\approx 4\sigma_{EM}$(AB;figure)
$\approx 8\sigma_{EM}$(GMM).
In the latter case, we are led to
$\sigma_{EM}$(TNSKM)
$\approx 2\sigma_{EM}$(AB;direct)
$\approx 4\sigma_{EM}$(GMM).
However, as mentioned in the text, we find
$\sigma_{EM}$(TNSKM) $\approx \sigma_{EM}$(GMM).
This puzzling situation can be avoided if the statement
$\sigma_{EM}$(GMM)$\approx 0.5\sigma_{EM}$(AB)
in Ref.\cite{mohanty}
is changed into the statement (in our notation)
$\sigma_{EM}$(GMM)$\approx 2\sigma_{EM}$(AB;direct).
In a private communication 
Dr. S. Mohanty has kindly informed us 
that he agrees with this change.}

\vspace{3mm}
In these circumstances
it is probably warranted to repeat
that, if one applies the same method (SNPA)
as used here to the calculation of 
the $n p \rightarrow \gamma d$ reaction cross section
\cite{wein95},
the results agree very well  
with the experimental data \cite{radiative}
for both the M1 and E1 contributions.
Furthermore, the calculation of the cross sections,
$\sigma(\nu_e d \rightarrow e^- pp)$ and 
$\sigma(\nu_x d \rightarrow \nu_x pn)$,
with the use of SNPA has been done by several groups
(with different degrees of sophistication),
and the results are known to be consistent among themselves
at the level of precision relevant to
the discussion of $\sigma_{EM}$ here \cite{kn94}.
Thus it is our belief
that the SNPA method used
in the present work has been well tested 
in both formal and numerical aspects.

\vspace{3mm} 
We now briefly discuss consequences 
of our calculation in the context
of obtaining information about the
neutrino magnetic moments.
In considering the SNO data,
we find it useful to follow 
the argument of GMM \cite{mohanty}.
The total ``NC" deuteron dissociation events
observed at SNO 
is the sum of the standard NC events
and those due to the $\mu_\nu$-interaction:
\begin{eqnarray}
N^{exp}=N_d T\Phi_{SUN}
[<\!\!\sigma_{NC}\!\!>+<\!\!\sigma_{EM}\!\!>]\,,
\end{eqnarray}
where $N_d$ is the number of target deuterons
and $T$ is the exposure time;
$\Phi_{SUN}$ is the total $^8$B solar neutrino flux. 
The empirical $^8$B neutrino flux, $\Phi_{SNO}$,
obtained at SNO assumes that the total dissociation
events arise solely from the standard NC interaction.
Thus
$\Phi_{SNO}\!=\!N^{exp}/[N_d T <\!\!\sigma_{NC}\!\!>]$,
which implies 
$\Phi_{SNO}=\Phi_{SUN}(1+\delta)$ with
$\delta\!\equiv
<\!\! \sigma_{EM}\!\!>/<\!\! \sigma_{NC}\!\!>$.
To extract information on $\mu_\nu$,
GMM employed $\sigma_{EM}=\sigma_{EM}$(GMM)
and the value of $\sigma_{NC}$ obtained by 
Nakamura et al.\cite{netal}, 
and furthermore they assumed 
$\Phi_{SUN}=\Phi_{SSM}$,
where $\Phi_{SSM}$ is the $^8$B solar neutrino flux 
predicted by the standard solar model \cite{BP01}.
GMM then arrived at 
$\Phi_{SNO}=\Phi_{SSM}(1+6.06\times 10^{14}{\tilde{\mu}}^2)$,
where ${\tilde{\mu}}$ is
the {\it operational} neutrino magnetic moment
defined by 
${\tilde{\mu}}^2=\mu_{21}^2+\mu_{22}^2+\mu_{23}^2$.
The use of 
$\Phi_{SNO}=(4.90\pm 0.37)
\times 10^6\,{\rm cm}^{-2}{\rm s}^{-1}$,
and
$\Phi_{SSM}=(5.87\pm 0.44)
\times 10^6\,{\rm cm}^{-2}{\rm s}^{-1}$
leads to 
${\tilde{\mu}}^2=(-2.76\pm 1.46)\times 10^{-16}$,
where the errors in $\Phi_{SNO}$ and $\Phi_{SSM}$
have been added quadratically.
Inflating the error up to 1.96$\sigma$,
GMM deduced the upper limit 
$|{\tilde{\mu}}|<3.71\times 10^{-9}\mu_B$
(95\% CL).
If we repeat a similar analysis
using $\sigma_{EM}$(TNSKM),
we arrive at
$\Phi_{SNO}=\Phi_{SSM}(1+5.46\times 10^{14}{\tilde{\mu}}^2)$.
We then should obtain practically the same conclusion 
on $|{\tilde{\mu}}|$ as in GMM.
This implies that, 
if the upper limit $\sim 10^{-10}\mu_B$
to $\mu_\nu$ is adopted,
the interpretation of the existing SNO data 
is not likely to be influenced 
by the possible existence of $\mu_\nu$.

\vspace{3mm}
Recently, an extensive global analysis of the totality of
the data from the Super-Kamiokande, 
SNO, and KamLAND has been carried out,
with the possible existence of $\mu_\nu$
taken into account \cite{nakahataSC}.
This analysis reports an upper limit
$\mu_{eff}<1.1\times 10^{-10}\mu_B$ (90 \% CL).
The value of $\sigma_{EM}$ 
described in our present communication
has been used as input in the global analysis
in Ref.\cite{nakahataSC}.

\vspace{3mm} 

\begin{figure}[h]
\caption{The neutrino deuteron disintegration cross 
sections, $\sigma_{NC}$ and $\sigma_{EM}$, 
as functions of $T_{NN}$. See the text for details. }
\includegraphics[width=8cm] {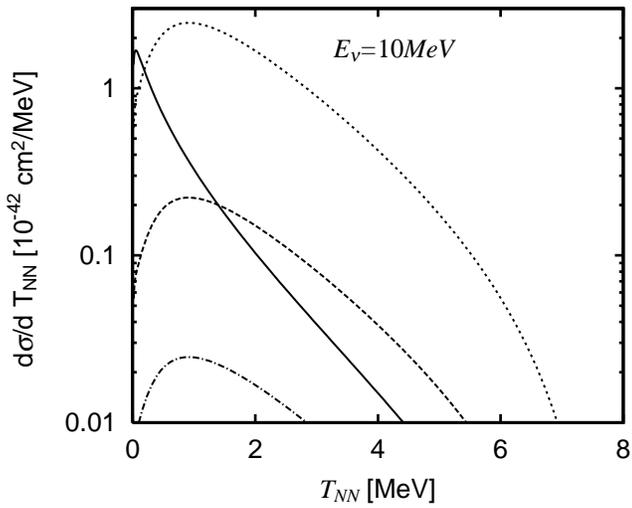} 
\label{fig-spec}
\end{figure}

It is known that a non-zero value of $\mu_\nu$ 
can affect the shape of the recoil electron spectrum
in electron-neutrino scattering. 
A similar effect is expected in 
the $\nu$-$d$ elastic scattering
(which is however at present practically
impossible to measure).
For the $\nu$-$d$ disintegration reaction,
the effect of the $\mu_\nu$ interaction is much smaller
than for the elastic scattering,
but it might be of some (academic) interest
to discuss to what extent one can distinguish
the NC and $\mu_\nu$-interaction contributions
to the neutron energy spectrum
in the $\nu$-$d$ breakup reaction. 
As an example, we consider 
the $T_{NN}$ dependences of 
$d\sigma_{EM}/dT_{NN}$ and 
$d\sigma_{NC}/dT_{NN}$,
where $T_{NN}$ is 
the final two-nucleon relative kinetic energy.
Fig.\ref{fig-spec} gives 
these differential cross sections
calculated for $E_\nu=10$ MeV,
and for $\mu_{eff} = 10^{-7}\mu_B, 
3\times 10^{-8}\mu_B$
and $10^{-8} \mu_B$.
These values of $\mu_{eff}$ are much larger 
than the upper limits known for $\nu_e$
and $\nu_\mu$,
but within the upper limit pertaining to $\nu_\tau$.
In Fig.\ref{fig-spec},
the NC contribution is shown by the solid line,
while the $\mu_\nu$ contribution
is given by the dotted, dashed and dot-dashed curves  
corresponding to $\mu_{eff} = 10^{-7}\mu_B, 
3\times 10^{-8}\mu_B$
and $10^{-8} \mu_B$, respectively.  
The NC contribution has a sharp peak 
in the low $T_{NN}$ energy region,
owing to the final $^1S_0$ state dominance
for the Gamow-Teller transition.  
By contrast, the $\mu_\nu$ contribution
exhibits a maximum away from threshold,
reflecting the fact that $\mu_\nu$-interaction
leads to the final P-wave states. 
As illustrated, $\mu_{eff}$
of the order of $10^{-7}\mu_B$ 
can affect the shape of the energy spectrum 
of the recoil nucleon 
in the $\nu$-$d$ breakup reaction. 

\vspace{3mm} 
To summarize, we have studied the effect of 
the neutrino magnetic moment
on the neutrino-deuteron breakup reaction,
using a method called 
the standard nuclear physics approach (SNPA),
which has been well tested for a number of
electroweak processes involving the deuteron.
The present calculation is free from 
the various approximations 
(mentioned earlier in the text)
that were used in the previous estimations.

\vskip 3mm \noindent
{\bf Acknowledgments}

We are grateful to M. Nakahata for a communication
that triggered the present study. 
We thank S. Mohanty and S. Nussinov 
for useful discussions.
This work was supported in part by the 
Japan Society for the Promotion of Science, 
Grant-in-Aid for Scientific Research (C) 15540275; 
by the 21st Century COE Program 
``Towards a new basic science: depth and synthesis";  
and by the US National Science Foundation,
Grant No. PHY-0140214.

\newpage
\begin{table}
\caption[]{
The cross section, $\sigma_{EM}$, for the
$\nu_x d \rightarrow \nu_x p n$ reaction
due to the $\mu_\nu$-interaction.
$\sigma_{EM}$ has been calculated 
for $\mu_{eff} = 10^{-10}\mu_B$,
and is given in units of ${\rm cm}^2$.}
 
\begin{tabular}{cccccccccc}
\hspace*{0.4cm}  $E_\nu$ & $\sigma_{EM}$ & \hspace*{0.4cm} $E_\nu$ & 
$\sigma_{EM}$ &
 \hspace*{0.4cm} $E_\nu$ & $\sigma_{EM}$ &
 \hspace*{0.4cm}   $E_\nu$ & $\sigma_{EM}$ & \hspace*{0.4cm} $E_\nu$ & 
$\sigma_{EM}$ 
\\ \hline
 2.4   & 2.16 (-53) &   6.0   & 7.55 (-49)  &   9.6   & 5.46 (-48)  &  13.2   & 
1.60 (-47)  &  16.8   & 3.28 (-47) \\
 2.6   & 2.81 (-52) &   6.2   & 8.86 (-49)  &   9.8   & 5.89 (-48)  &  13.4   & 
1.67 (-47)  &  17.0   & 3.39 (-47) \\
 2.8   & 1.23 (-51) &   6.4   & 1.03 (-48)  &  10.0   & 6.33 (-48)  &  13.6   & 
1.75 (-47)  &  17.2   & 3.51 (-47) \\
 3.0   & 3.51 (-51) &   6.6   & 1.19 (-48)  &  10.2   & 6.79 (-48)  &  13.8   & 
1.83 (-47)  &  17.4   & 3.62 (-47) \\
 3.2   & 7.88 (-51) &   6.8   & 1.36 (-48)  &  10.4   & 7.26 (-48)  &  14.0   & 
1.91 (-47)  &  17.6   & 3.74 (-47) \\
 3.4   & 1.52 (-50) &   7.0   & 1.55 (-48)  &  10.6   & 7.76 (-48)  &  14.2   & 
2.00 (-47)  &  17.8   & 3.86 (-47) \\
 3.6   & 2.63 (-50) &   7.2   & 1.75 (-48)  &  10.8   & 8.28 (-48)  &  14.4   & 
2.09 (-47)  &  18.0   & 3.98 (-47) \\
 3.8   & 4.21 (-50) &   7.4   & 1.97 (-48)  &  11.0   & 8.81 (-48)  &  14.6   & 
2.17 (-47)  &  18.2   & 4.11 (-47) \\
 4.0   & 6.35 (-50) &   7.6   & 2.20 (-48)  &  11.2   & 9.36 (-48)  &  14.8   & 
2.27 (-47)  &  18.4   & 4.23 (-47) \\
 4.2   & 9.13 (-50) &   7.8   & 2.45 (-48)  &  11.4   & 9.94 (-48)  &  15.0   & 
2.36 (-47)  &  18.6   & 4.36 (-47) \\
 4.4   & 1.26 (-49) &   8.0   & 2.72 (-48)  &  11.6   & 1.05 (-47)  &  15.2   & 
2.45 (-47)  &  18.8   & 4.49 (-47) \\
 4.6   & 1.70 (-49) &   8.2   & 3.00 (-48)  &  11.8   & 1.11 (-47)  &  15.4   & 
2.55 (-47)  &  19.0   & 4.62 (-47) \\
 4.8   & 2.21 (-49) &   8.4   & 3.30 (-48)  &  12.0   & 1.18 (-47)  &  15.6   & 
2.65 (-47)  &  19.2   & 4.76 (-47) \\
 5.0   & 2.83 (-49) &   8.6   & 3.62 (-48)  &  12.2   & 1.24 (-47)  &  15.8   & 
2.75 (-47)  &  19.4   & 4.89 (-47) \\
 5.2   & 3.54 (-49) &   8.8   & 3.95 (-48)  &  12.4   & 1.31 (-47)  &  16.0   & 
2.85 (-47)  &  19.6   & 5.03 (-47) \\
 5.4   & 4.37 (-49) &   9.0   & 4.30 (-48)  &  12.6   & 1.38 (-47)  &  16.2   & 
2.95 (-47)  &  19.8   & 5.17 (-47) \\
 5.6   & 5.31 (-49) &   9.2   & 4.67 (-48)  &  12.8   & 1.45 (-47)  &  16.4   & 
3.06 (-47)  &  20.0   & 5.31 (-47) \\
 5.8   & 6.36 (-49) &   9.4   & 5.06 (-48)  &  13.0   & 1.52 (-47)  &  16.6   & 
3.17 (-47)  & &
\end{tabular}
\end{table}


\begin{thebibliography}{99}

\bibitem{sno}
    Q. R. Ahmad et al., 
    Phys. Rev. Lett. {\bf 87} (2001) 071301 ;
    {\it ibid} {\bf 89} (2002) 011301 ;
    {\it ibid} {\bf 89} (2002) 011302.

\bibitem{snox}
   S. N. Ahmed et al., Phys. Rev. Lett. {\bf 92} (2004) 181301.

\bibitem{nsgk}
    S. Nakamura, T. Sato, V. Gudkov and K. Kubodera,
    Phys. Rev. C {\bf63} (2001) 034617.

\bibitem{netal}
    S. Nakamura, T. Sato, S. Ando, T.-S. Park, 
    F. Myhrer, V. Gudkov and K. Kubodera, 
    Nucl. Phys. A {\bf 707} (2002) 561.

\bibitem{butler1}
    M. Butler and J.-W. Chen, 
    Nucl. Phys. {\bf A675} (2000) 575.

\bibitem{butler2}
    M. Butler, J.-W. Chen and X. Kong, 
    Phys. Rev. {\bf C63} (2001) 035501.

\bibitem{ando}
    S. Ando, Y. H. Song, T. -S. Park,
    H. W. Fearing and  K. Kubodera, 
    Phys. Lett. B {\bf 555} (2003) 49.

\bibitem{Parketal03}
    T.-S. Park et al., 
    Phys. Rev. C {\bf 67} (2003) 055206.

\bibitem{ab}
    E. Kh. Akhmedov and V. V. Berezin, 
    Z. Phys. C {\bf 54} (1992) 661.

\bibitem{mohanty}
    J.A. Grifols, E. Masso and S. Mohanty,  
    Phys. Lett. B {\bf 587} (2004) 184;
    hep-ph/0401144.

\bibitem{wein95}
    T. Sato, T. Niwa and H. Ohtsubo,
    in {\it Proceedings of the IVth International 
    Symposium on Weak and Electromagnetic Interactions
    in Nuclei}, edited by H. Ejiri, T. Kishimoto and 
    T. Sato (World Scientific, Singapore, 1995), p. 488.

\bibitem{raf} 
   G.G. Raffelt,   
   {\it Stars as Laboratories for Fundamental Physics},
   The University of Chicago Press, Chicago \& London
   (1996)

\bibitem{fy} 
   M. Fukugita and T. Yanagida, 
   {\it Physics of Neutrinos and Applications to 
      Astrophysics}, Springer, Berlin (2003).

\bibitem{anl}
    R.B. Wiringa, V.G.J. Stoks, R. Schiavilla, 
    Phys. Rev. C {\bf 51} (1995) 38.

\bibitem{MUNU}
     Z. Daraktchieva et al.(MUNU Collaboration),
     Phys. Lett. B {\bf 564} (2003) 190.

\bibitem{TEXONO}
     H. B. Li et al.(TEXONO Collaboration), 
     Phys. Rev. Lett. {\bf 90} (2003) 131802.

\bibitem{LSND}
     L.B. Auerbach et al.(LSND Collaboration),
     Phys. Rev. D {\bf63} (2001) 112001.

\bibitem{DONUT}
    R. Schwienhorst et al.(DONUT Collaboration), 
    Phys. Lett. B {\bf 513} (2001) 23.

\bibitem{beacom}
    J. F. Beacom and P. Vogel, 
    Phys. Rev. Lett. {\bf 83} (1999) 5222.

\bibitem{radiative}
   T.S. Suzuki, Y. Nagai, T. Shima, T. Kikuchi, H. Sato,
   T. Kii, and M. Igashira,
   Astrophys. J. Lett. {\bf 439} (1995) L59;
   Y. Birenbaum, S. Kahane, and R. Moreh,
   Phys. Rev. C, {\bf 32} (1985) 1825;
   R. Bernabei {\it et al.}, 
   Phys. Rev. Lett. {\bf 57} (1986) 1542.



\bibitem{kn94}
   K. Kubodera and S. Nozawa,
   Int. J. Mod. Phys. E, {\bf 3} (1994) 101;
   K. Kubodera, nucl-th/0404027.

\bibitem{BP01}
   J.N. Bahcall, M.H. Pinsonneault and S. Basu,
   Astrophys. J. {\bf 555} (2001) 990.

\bibitem{nakahataSC} 
    D. W. Liu et al. (Super-Kamiokande Collaboration),
    Phys. Rev. Lett., {\bf 93} (2004) 021802.


\end{thebibliography}
\end{document}